\documentclass[aps,showpacs,amsmath,amssymb,nofootinbib]{revtex4}
\usepackage{graphicx}
\usepackage{dcolumn}
\usepackage{bm}
\usepackage{epsfig}
\usepackage{multirow}

\begin{document}
\title{Quasinormal modes of the Schwarzschild black hole with a deficit solid angle and quintessence-like
	matter: Improved asymptotic iteration method}

\author{L. A. L\'opez$^1$}
\email{lalopez@uaeh.edu.mx}
\author{Omar Pedraza$^1$}
\email{omarp@uaeh.edu.mx}
\author{Roberto Arceo$^2$}
\email{roberto.arceo@unach.mx}
\author{V. E. Ceron$^1$}
\email{vceron@uaeh.edu.mx}

\affiliation{$^1$ \'Area Acad\'emica de Matem\'aticas y F\'isica, UAEH, 
Carretera Pachuca-Tulancingo Km. 4.5, C. P. 42184, Mineral de la Reforma, Hidalgo, M\'exico.}
\affiliation{$^2$Facultad de Ciencias en F\'isica y Matem\'aticas, Universidad Aut\'onoma de Chiapas,
C. P. 29050, Tuxtla Guti\'errez, Chiapas, M\'exico}



\begin{abstract}
We study the quasinormal modes (QNM) for scalar,  and electromagnetic perturbations in the Schwarzschild black hole with a deficit solid angle and quintessence-like matter. Using the sixth--order WKB approximation and the improved asymptotic iteration method (AIM) we can determine the dependence of the quasinormal modes on the parameters of the black hole and the parameters on the test fields. The values of the real part and imaginary parts of the quasi--normal modes increase with the decrease of the values of the deficit solid angle and density of quintessence-like matter. The quasinormal modes gotten by these two methods are in good agreement. Using the finite difference method, we obtain the time evolution profile of such perturbations in this Black Hole. 
\end{abstract}

\maketitle
\section{Introduction}

One state resulting from the disturbance (scalar or electromagnetic perturbations) of a black hole (BH)  is an oscillation with complex frequencies called quasinormal modes (QNM). The QNM is related to the parameters that describe the BH as mass, charge, angular moment, and others. But also, we can analyze the frequency of oscillation and the damping of the oscillation through the real part and imaginary part of the frequencies of the QNM respectively. Another important contribution of the QNM is the analysis of the stability of the BHs and their gravitational radiation.

The study of QNM has been carried out for different solutions of black holes that represent isolated solutions or in vacuum for example; QNM of Schwarzschild \cite{Iyer:1986nq}, Reissner–Nordström, Hayward \cite{Lin:2013}, and others. Also, different authors have investigated QNM in non-linear electrodynamics (NLED); QNM of Bardeen \cite{Fernando:2012yw}, Einstein-Born-Infeld \cite{Lee:2020iau}, and  Generic-class \cite{Lopez:2022uie} to mention some investigations. 

Now, if we consider that the black holes' coexistence with other types of matter or energy is significant, and according to different observations, dark energy is distributed throughout the universe and is the cause of its expansion. Then, the study of black holes surrounded by dark energy is a topic of current interest.

There are different models as candidates for dark energy and the difference between the models is the magnitude of the state parameter that represents the ratio of pressure to the energy density of dark energy. One of the candidates is the quintessence which differs with the cosmological constant in the magnitude of the state parameter.

The studies of QNM of BHs surrounded by the quintessence have been developed for different scenarios as Hayward \cite{Pedraza:2021hzw}, Bardeen, Reissner–Nordström \cite{Saleh:2011zz} and others black holes, all surrounded by quintessence (using the Kiselev model \cite{Kiselev:2002dx}). 

In the special case of the Schwarzschild solution, the QNM have been studied in different scenarios considering the influence of dark energy. for example; in \cite{Zhang:2006ij} and \cite{Chen:2005qh} studied the QNM of Schwarzschild surrounded by quintessence, also QNM of a Schwarzschild black hole with a deficit angle and quintessence-like matter are investigated in \cite{Yu:2022yyv}, in \cite{Hamil:2024ppj} studied the QNM of noncommutative Schwarzschild black hole surrounded by quintessence.

In the works above mentioned on the study of QNM,  different numerical methods have been applied being the WKB method the most used. When you want to perform the analysis of QNM, this lead to a wave equation with a specific effective potential, depending on the characteristics of the effective potential some numerical methods are more suitable. 

The asymptotic iteration method (AIM) proposed in \cite{HakanCiftci_2003} to obtain the eigenvalues and eigenfunctions of linear differential equations second-order homogeneous, has shown is a method can be a technique efficient and accurate in calculating the frequencies of QNM for a wide variety of black holes, for example in \cite{Cho:2011sf}  the QNM of  Schwarzschild, Schwarzschild anti-de Sitter or de Sitter, Reissner–Nordström and Kerr are analyzed with AIM. When you want to perform the analysis of QNM, this lead to a wave equation with a specific effective potential, depending on the characteristics of the effective potential some numerical methods are more suitable.

This paper is devoted to the study of the QNM and the time evolution of massless scalar and electromagnetic perturbations on Schwarzschild black hole with a deficit solid angle and quintessence-like matter. The wave-like perturbation equation, with an effective potential, can be solved to calculate the QNM numerically by two methods: the WKB approach (see \cite{Xi:2008ce,Xi:2010pv,Wang:2012vvx}) and AIM (to demonstrate the viability of the method to obtain quasinormal modes). The paper is organized as follows: Section \ref{sec.1} we present the Black Hole considered in the present work and we briefly discuss the event horizons. In Sec. \ref{sec.2} we describe the scalar and electromagnetic perturbations of a Black Hole. We study the behavior of effective potential for different perturbations considering the special cases when the quintessence state parameter takes the values -1/2 and -2/3. In Sec. \ref{sec.3} we present the results for quasinormal frequencies (QNFs) and in Sec.\ref{sec.4} we present the time domain analysis. Finally, our conclusions are in Sec. \ref{sec.5}.

\section{A Schwarzschild black hole with quintessence-like matter and a deficit solid angle}\label{sec.1}
Barriola and Vilenkin \cite{Barriola:1989hx} proposed static and spherical symmetric solution that describe Schwarzschild Black Hole with a deficit solid angle and quintessence-like matter. This solution is given by
\begin{equation}\label{mfa}
ds^2=f(r)dt^2-\frac{dr^2}{f(r)}-r^2d\theta^2-r^2\sin^2\theta d\phi^2\,,
\end{equation}
where 
\begin{equation}\label{ec.rfc}
f(r)=1-\frac{2M}{r}-\epsilon^2+\frac{\rho_0}{3\omega_0 r^{3\omega_0+1}}\,.
\end{equation}
In which $\epsilon$ is the deficit solid angle parameter, $\rho_0$ is the density of quintessence-like matter, $\omega_0$ is the quintessence state parameter and satisfies $-1<\omega_0<-1/3$ and $M$ is the black hole mass. When $\epsilon=\rho_{0}=0$ the metric function $f(r)$ is reduced to the solution of Schwarzschild. 

The horizons of Schwarzschild with quintessence-like matter and a deficit solid angle (Schwdm) are determined by the positive roots of the equation $f(r) = 0$, this condition leads to the polynomial
\begin{equation}\label{ec.hwq}
3\omega_0\left(\epsilon^2-1\right) r^{3\omega_0+1}+6M\omega_0 r^{3\omega_0}-\rho_0=0\,.
\end{equation}

\begin{figure}[!h]
	\centering
	\includegraphics[scale=0.93]{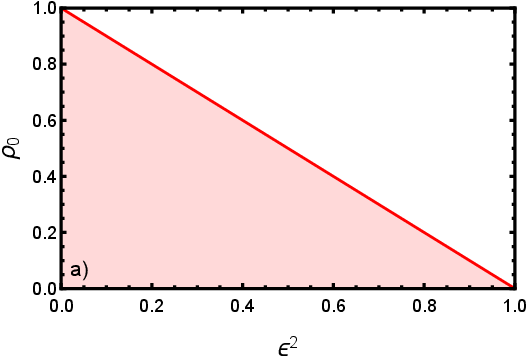}
	\includegraphics[scale=0.95]{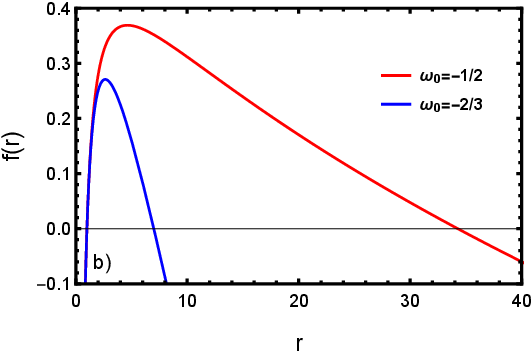}
	\caption{a) The parametric region where the Schwdm BH has two horizons, $r_+\leq r_c$ according to $\omega_0$ values. The region is bounded by the line $(\rho_0=1-\epsilon^2)$, $\rho_0=0$ and $\epsilon=0$ (the region shaded). b) The graph shows the metric function (\ref{ec.rfc}) as function of $r$ for different values of $\omega_0$.  Here $\rho_{0}=0.2$ and $\epsilon^2=0.2$.}
	\label{g1}
\end{figure}

The number of horizons depends entirely on the choice of the values of parameters $\epsilon$, $\omega_0$, $M$ and $\rho_0$. However, in an appropriated parametric region, the solution has two horizons: the event horizon $r_+$ and the cosmological horizon $r_c$. Using the ideas of \cite{Liu:2020evp}, here hereafter, we fix the  event horizon in  $r_+=1$. To ensure $f(r_+)=0$, it is required that the mass parameter can be expressed as
\begin{equation}\label{ec.m}
M=\frac{3\omega_0\left(1-\epsilon^2\right)+\rho_0}{6\omega_0}\,.
\end{equation} 
The region parametric in which the Schwdm BH has event horizon $r_+$ and cosmological horizon $r_c$ are allowed, can be determined by $f'(r_+)>0$. This condition leads to the next expression; 
\begin{equation}\label{ecu.rre}
\rho_0<1-\epsilon^2\,.	
\end{equation}
The region plot of the allowed parameter region is given in Fig. \ref{g1}(a). For all values of ($\epsilon^2,\rho_0$) in region shaded, there are two real positive roots of Eq. (\ref{ec.hwq}), $r_+=1$ and $r_c$, which satisfy $r_+\leq r_c$. On the other hand, for values of ($\epsilon^2,\rho_0=1-\epsilon^2$) corresponding to the line represents extremal Schwdm BH, where $r_+=r_c=1$. In the Fig.  \ref{g1}(a) we observe that the parametric region is independent of $\omega_0$. From Fig. \ref{g1}(b), we plot the metric function $f(r)$ for different values of $\omega_0$ and we can conclude that the quintessence horizon is very large for $\omega_0\to-1/3$.     

Now inserting $M$ (\ref{ec.m}) into (\ref{ec.rfc}), we can write the metric function as
\begin{equation}
f(r)=1-\epsilon^2-\frac{1-\epsilon^2+\frac{\rho_0}{3\omega_0}}{r}
+\frac{\rho_0}{3\omega_0 r^{3\omega_0+1}}\,.
\end{equation}
For the particular case $\omega_0=-2/3$, the cosmological and event horizons are given by
\begin{equation}
	r_+=1,\quad r_c=\frac{2\left(1-\epsilon^2\right)-\rho_0}{\rho_0}\,.
\end{equation} 
And for $\omega_0=-1/2$, we get
\begin{equation}
	r_+=1,\quad r_c=\frac{\left(3\left[1-\epsilon^2\right]-2\rho_0\right)}{2\rho_0}\left(1+
	\frac{3\left[1-\epsilon^2\right]-2\rho_0+\sqrt{3\left(3\left[1-\epsilon^2\right]-2\rho_0\right)\left(1-\epsilon^2+2\rho_0\right)}}{4\rho_0}	
	\right)\,.
\end{equation} 
When the event and cosmological horizons coincide with each other, the Schwdm BH is an extremal one ($\rho_0=1-\epsilon^2$), thus $r_+=r_c=1$.

To conclude this section, it is worth mentioning that if $r_+\neq1$ then expression (\ref{ecu.rre}) takes the following form
\begin{equation}\label{ec.rvse}
\rho_0<\left(1-\epsilon^2\right)\left[
\frac{\left(1-\epsilon^2\right)\left(3|\omega_0|-1\right)}{6M|\omega_0|}
\right]^{3|\omega_0|-1}\,,
\end{equation}
for further discussion, see the Ref. \cite{Yu:2022yyv}. Which depends explicitly on $\omega_0$, so comparing the different parametric regions, it is possible to mention that the parametric region for $\omega_0=-2/3$ is much smaller than the parametric region for $\omega_0=-1/2$. It is worth mentioning that when we fix $r_+=1$, the parametric region ($\epsilon^2,\rho_0$) (see Eq. (\ref{ecu.rre}) and Fig. \ref{g1}) is independent of $\omega_0$, which facilitates the analysis of quasinormal modes.

\section{Scalar and Electromagnetic perturbations}
\label{sec.2}
In this section, we briefly show the behavior of scalar and electromagnetic perturbations in a Black Hole with a deficit solid angle and quintessence-like matter, following the Refs. \cite{Xi:2010pv,Wang:2012vvx}. The propagation of a scalar field in a curved background is described by the Klein-Gordon equation
\begin{equation}\label{ec.scalar}
\frac{1}{\sqrt{-g}}\partial_{\mu}\left(\sqrt{-g}\,g^{\mu\nu}\partial_{\nu}\Psi\right)=0\,.
\end{equation}
We separate variables by setting 
\begin{equation}\label{sol}
	\Psi(t,r,\theta,\phi)=\frac{1}{r}\phi(r)Y_{lm}(\theta,\phi)e^{i\omega t}\,,
\end{equation}
where $Y_{lm}(\theta,\phi)$ are the spherical harmonics. Substituting Eq. (\ref{sol}) into (\ref{ec.scalar}), we get the next radial equation  
\begin{equation}\label{ecu.cs}
	\left[\frac{d^2}{dr_*^2}+\omega^2-V_s(r)
	\right]\phi(r)=0\,,
\end{equation}
where $r_*$ is the tortoise coordinate
\begin{equation}
	dr_*=\frac{dr}{f(r)}\,,
\end{equation}
and $V_s(r)$ is the effective potential
\begin{equation}\label{ecu.vs}
	V_s(r)=f(r)\left[
\frac{l(l+1)}{r^2}+\frac{f'(r)}{r}
\right]\,,
\end{equation}
whereas, the electromagnetic field in curved space follows the next equation
\begin{equation}\label{ec.el}
	\frac{1}{\sqrt{-g}}\partial_{\mu}\left(\sqrt{-g}\,F_{\beta\gamma}g^{\beta\nu}g^{\gamma\mu}\right)=0\,,
\end{equation}
where $F_{\beta\gamma}=\partial_{\beta}A_{\gamma}-\partial_{\gamma}A_{\beta}$ and the vector potential $A_{\mu}$ can be expressed as 
\begin{equation}\label{ecu.va}
	A_0=A_1=A_2=0,\quad
	A_3=\phi_e(r)\sin(\omega t)\sin\theta\frac{dP_l(\cos\theta)}{d\theta}\,.
\end{equation} 
After separation of variables, the radial parts of the electromagnetic field perturbation equation takes the form similar to (\ref{ecu.cs}), but the effective potential is given by 
\begin{equation}\label{veltf}
	V_e(r)=f(r)\frac{l(l+1)}{r^2}\,.
\end{equation}
From Eqs. (\ref{ecu.vs}) and (\ref{veltf}), the scalar and electromagnetic perturbations can be described by a Schr\"odinger-like equation with potential    
\begin{equation}\label{ec.ept}
V(r)=f(r)\left[
\frac{l(l+1)}{r^2}+(1-s)\frac{f'(r)}{r}
\right]\,,
\end{equation}
where $l$ is the multipole number ($\l\geq s$), $s$ is the spin of the perturbative field: $s=0$ corresponding to scalar perturbation and $s=1$ to electromagnetic perturbation. The effective potential (\ref{ec.ept}) has asymptotic value $V(r)\approx \rho_0^2\left(
s-1\right)\left(1+s+3\omega_0\right)/\left(9\omega_0^2r^{6\omega_0+4}\right)$ when $r\to\infty$. From this behavior asymptotic for the effective potential, we can see that if $\omega_0<-2/3$, $V\to\infty$ when $r\to\infty$. Thus we only study the case where the state parameter which has the range $-2/3\leq\omega_0<-1/3$. For the specific case $\omega_0=-2/3$, the scalar perturbation, $V(r)\approx \rho_0^2/4$. 

\begin{figure}[!h]
	\centering
	\includegraphics[scale=0.95]{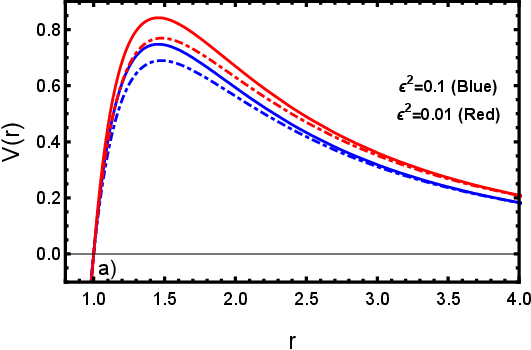}
	\includegraphics[scale=0.95]{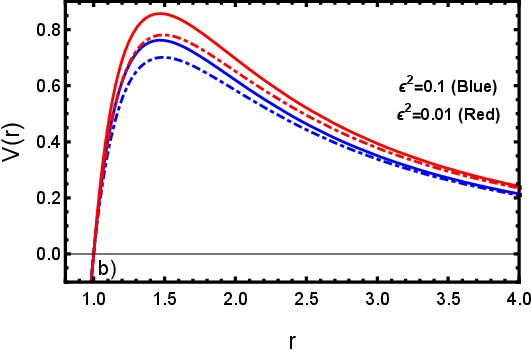}
	\caption{a) The behavior of the effective potential for $\omega_q=-2/3$ is shown for various values of $\epsilon^2$, with $\rho_0=0.1$ and $l=2$. b) The behavior of the effective potential for $\omega_q=-1/2$ is shown for various values of $\epsilon^2$, with $\rho_0=0.1$ and $l=2$. In both figures, the effective potential of scalar (solid) and electromagnetic (dot dashed)}
	\label{g2}
\end{figure} 

\begin{figure}[!h]
	\centering
	\includegraphics[scale=0.95]{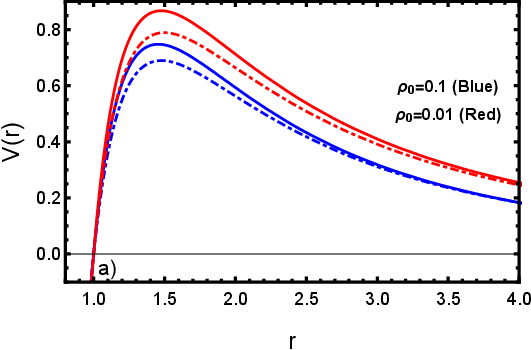}
	\includegraphics[scale=0.95]{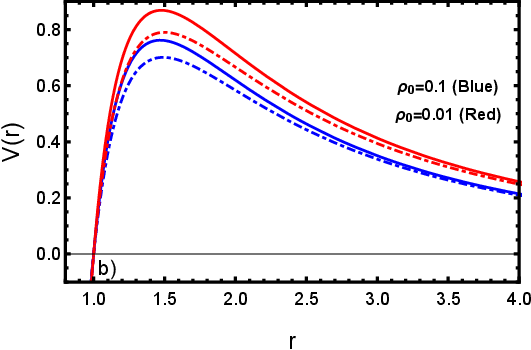}
	\caption{a) The behavior of the effective potential for $\omega_q=-2/3$ is shown for various values of $\rho_0$, with $\epsilon^2=0.1$ and $l=2$. b) The behavior of the effective potential for $\omega_q=-1/2$ is shown for various values of $\rho_0$, with $\epsilon^2=0.1$ and $l=2$. In both figures, the effective potential of scalar (solid) and  electromagnetic (dot dashed)}
	\label{g3}
\end{figure} 

The behavior of $V(r)$ for different values of $\epsilon^2$ is described in Fig. \ref{g2} a) and b) for $\omega_q=-2/3$ and $\omega_q=-1/2$ respectively,  while the behavior of effective potential for different values of $\rho_0$ is described in Fig. \ref{g3} for different values of state parameter. Furthermore, the form of the effective potential is very similar to that obtained in the quintessence case for scalar and fermionic perturbations \cite{ALBADAWI}.

It is worth mentioning that the effective potential is positive definite for $r_*\in\left[-\infty,\infty\right]$ and have a potential barrier near the event horizon for very small values of the state parameter and deficit solid angle parameter.

From Figs. \ref{g2} and \ref{g3} we can mention that in both cases $V_{\text{elec}}(r)<V_{\text{sc}}(r)$.

\begin{figure}[!h]
	\centering
	\includegraphics[scale=0.95]{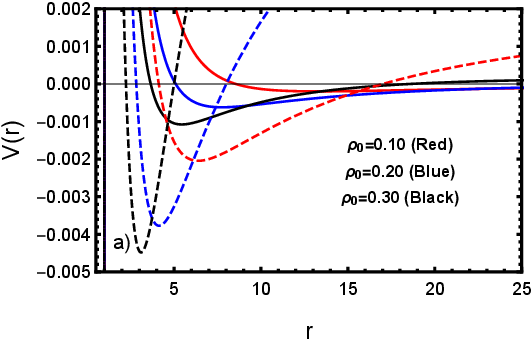}
	\includegraphics[scale=0.95]{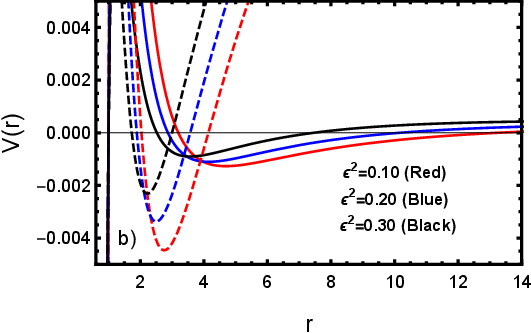}
	\caption{The effective potential of scalar field with $l=0$ mode. The a) and b) plots correspond to $\epsilon^2=0.1$ and  $\rho=0.35$ respectively. In both figures, the effective potential for $\omega_0=-1/2$ is represented by the solid line and by the dotted line for $\omega_0=-2/3$.}
	\label{g4}
\end{figure} 

If we study the effective potential for scalar field with $l=0$ mode, we can see that $V(r)$ can take negative values between r$_+$ and $r_c$ (see Figs. \ref{g4}(a) and \ref{g4}(b)). This potential well is the key point for the occurrence of instability. However, the negative effective potential does not guarantee that the imaginary part of the quasinormal frequencies takes positive values i.e. the existence of a negative potential well can be viewed as the necessary but not sufficient condition for instability \cite{PhysRevD.86.024028}.

Also we can see in Fig. \ref{g4}(a) that for increasing values of $\rho_0$, the depth of the potential well increases, while in Fig. \ref{g4}(b) shows that for increasing values of $\epsilon^2$, the depth of the potential well decreases. The stability or instability analysis of scalar perturbations for the case $l=0$, is discussed in Sec. \ref{sec.4}.  

The Eq. (\ref{ecu.cs}) exhibits the following behavior near the horizons
\begin{equation}\label{key4}
\phi_s\approx\begin{cases}
e^{-i\omega r_*}, & r_*\to-\infty\,(r\to r_+)\\
e^{i\omega r_*}, & r_*\to\infty\,(r\to r_c)
	\end{cases}\,,
\end{equation}

the asymptotic solutions (\ref{key4}) imply that the wave at the event horizon is purely incoming, whereas at the cosmological horizon is purely outgoing. For the boundary condition mentioned above, the wave equation (\ref{ecu.cs}) can be solved in order to obtain the spectrum of the quasi normal frequencies.  

\section{Qnm: Improved asymptotic iteration method}
\label{sec.3}

It is well known that in the literature, there are some numerical and semianalytical methods to calculate QNM. The method suggested by Mashhoon \cite{BLOME1984231}, where the potential barrier in the Schrödinger-like equation is replaced by a potential barrier for which we know the analytic solution. This method gives quite accurate results for the regime of high multipole numbers.  Besides the method by Leaver \cite{Leaver:1985ax}, typically by applying a Frobenius series solution approach, leading to continued fractions for QNM.  Another technique is the WKB approximation (this approach has been extended to the sixth order, see \cite{konoplya2003quasinormal}), this approach gives more accuracy and can be carried to a higher order. However, the WKB approach does not give reliable frequencies for $n\geq l$. As mentioned, these techniques have some limitations. On the other hand, another new method that has been developed that may be more efficient in some cases is the asymptotic iteration method.

The AIM has been used to calculate the QNM of black holes in the asymptotic dS spacetimes \cite{Cho:2011sf, Cho:2009cj} 
To apply AIM, we need to introducing the transformation of coordinates $\xi=1/r$. In the new coordinate $\xi$, the equation (\ref{ecu.cs}), which leads to the following expression
\begin{equation}\label{ev.epsee}
\left[\frac{d^2}{d\xi^2}+\frac{p'}{p}\frac{d}{d\xi}+
\frac{\omega^2}{p^2}-\frac{
l\left(l+1\right)+\left(1-s^2\right)\left(
1-\epsilon^2-\frac{p}{\xi^2}
\right)-\left(1-s\right)\rho_0\xi^{3\omega_0+1}	
}{p}\right]\phi=0
\,,
\end{equation}
where the prime denotes derivative with respect $\xi$, and $p(\xi)$ is defined as
\begin{equation}
p(\xi)=\left(1-\epsilon^2\right)\xi^2
-\frac{3\omega_0\left(1-\epsilon^2\right)+\rho_0}{3\omega_0}\xi^{3}
+\frac{\rho_0}{3\omega_0}\xi^{3(\omega_0+1)}
\,.
\end{equation}
To accommodate the outgoing wave boundary condition $\phi\to e^{i\omega r_*}$ in terms of $\xi$, we define
\begin{equation}\label{key0}
e^{i\omega r_*}=e^{-i\omega\int \frac{d\xi}{p(\xi)}}\,.
\end{equation}
The expression (\ref{key0}) scale out the divergence behavior at the cosmological horizon. So, by implement the boundary conditions in the AIM, we redefine the function $\phi$ in terms of the new function $\chi$. So we can write 
\begin{equation}\label{ec.bcrm}
\phi\left(\xi\right)=e^{-i\omega\int \frac{d\xi}{p(\xi)}}\left(\xi-\xi_+\right)^{-i\omega/\kappa_+}\chi\left(\xi\right)\,.
\end{equation}
Considering (\ref{ec.bcrm}),  we can scale out the divergent behavior at the event horizon. Here $\kappa_+$ is the surface gravity at the horizon defined as;
\begin{equation}\label{key}
\kappa_+=\left.\frac{1}{2}\frac{d}{dr}f(r)\right|_{r\to r_+}=-\left.\frac{\xi^2}{2}\frac{d}{d\xi}f(\xi)\right|_{\xi\to \xi_+}\,.
\end{equation}
By combining equations (\ref{ev.epsee}) and (\ref{ec.bcrm}), we obtain the following differential equation for $\chi(\xi)$
\begin{equation}\label{key1}
\frac{d^2\chi\left(\xi\right)}{d\xi^2}
=\lambda_0\left(\xi\right)
\frac{d\chi\left(\xi\right)}{d\xi}
+s_0\left(\xi\right)\chi\left(\xi\right)\,,
\end{equation}
where
\begin{eqnarray}
\lambda_0\left(\xi\right)&=&
\frac{2i\omega}{\kappa_+\left(\xi-\xi_+\right)}-\frac{p'-2i\omega}{p}
\label{key2a} \,,\\
s_0\left(\xi\right)&=&
\frac{l\left(l+1\right)+\left(1-s\right)\left(
	1-\epsilon^2-\frac{p}{\xi^2}
	-\rho_0\xi^{3\omega_0+1}\right)}{p}
-\frac{i\omega\left(\kappa_++i\omega\right)}{\kappa^2_+\left(\xi-\xi_+\right)^2}
+
\frac{i\omega\left(p'-2i\omega\right)}{p\kappa_+\left(\xi-\xi_+\right)}
\,.\label{key2b}
\end{eqnarray}
 
Now following the improved AIM procedure, the expressions (\ref{key2a}) and (\ref{key2b}) are used in the relations (\ref{ec.ln}) and (\ref{ec.sn}) (see Appendix) to calculate the quantities that appear in the quantization condition (\ref{ec.qz1}). The stable roots of condition (\ref{ec.qz1}) are the QNM of the scalar and electromagnetic test fields. We use Mathematica Software to perform all the calculations.

We present in Tables \ref{s0e}, \ref{s0r}, \ref{s1e} and \ref{s1r} results for scalar and electromagnetical perturbation respectively. We compared our results obtained using AIM (after of fifteen iterations) with WKB  to sixth-order. From Tables \ref{s0e}-\ref{s1r}, we can see that between the WKB method and AIM, there are differences in the values of QNM at $n = l = 0$, (it is well known that WKB approximation does not give reliable frequencies for $n\geq l$ ) and the case for higher $l$ the differences are smaller between both methods.

\begin{table}[th]
\begin{center}
\begin{tabular}{l c c c c c c c c}
\hline
\hline
\multicolumn{9}{c}{Scalar perturbations}\\ \hline
&&\multicolumn{3}{c}{$\omega_0=-2/3$}
&\multicolumn{4}{c}{$\omega_0=-1/2$}\\
\hline
&&\multicolumn{1}{c}{WKB}
&&\multicolumn{1}{c}{AIM}
&&\multicolumn{1}{c}{WKB}
&&\multicolumn{1}{c}{AIM}\\ 
	$\epsilon^2$&\multicolumn{8}{c}{$l=0$}
\\
0.01
&&0.187043-0.191982$i$	
&&0.186947-0.195326$i$	
&&0.190396-0.184349$i$
&&0.190432-0.189805$i$
\\
0.10
&&0.166671-0.174322$i$	
&&0.167064-0.176485$i$	
&&0.170528-0.166209$i$
&&0.170589-0.170975$i$
\\		
$\epsilon^2$&\multicolumn{8}{c}{$l=1$}\\
0.01
&&0.536367-0.175563$i$	
&&0.536404-0.175406$i$	
&&0.542744-0.174016$i$
&&0.542784-0.173865$i$
\\
0.10
&&0.504477-0.15763$i$	
&&0.504505-0.157515$i$	
&&0.510936-0.156189$i$
&&0.510967-0.156078$i$
\\
&\multicolumn{8}{c}{$l=3$}\\
0.01
&&1.252810-0.172128$i$	
&&1.252810-0.172127$i$	
&&1.263260-0.171518$i$
&&1.263260-0.171517$i$
\\
0.10
&&1.184560-0.154697$i$	
&&1.184560-0.154696$i$	
&&1.195430-0.154106$i$
&&1.195430-0.154105$i$
\\
&\multicolumn{8}{c}{$l=5$}\\
0.01
&&1.969140-0.171672$i$	
&&1.969140-0.171672$i$	
&&1.984550-0.171193$i$
&&1.984550-0.171193$i$
\\
0.10
&&1.863150-0.154311$i$	
&&1.863150-0.154311$i$	
&&1.879280-0.153837$i$
&&1.879280-0.153837$i$
\\
&\multicolumn{8}{c}{$l=7$}\\
0.01
&&2.685400-0.171528$i$	
&&2.685400-0.171528$i$	
&&2.705960-0.171091$i$
&&2.705960-0.171091$i$
\\
0.10
&&2.541410-0.154190$i$	
&&2.541410-0.154190$i$	
&&2.562970-0.153753$i$
&&2.562970-0.153753$i$
\\
\hline
\hline
\end{tabular}
\caption{Quasinormal frequencies for the Scalar perturbations for several values of the parameter $\epsilon^2$, with $n=0$ and $\rho_0=0.1$.}
\label{s0e}
\end{center}
\end{table}

\begin{table}[th]
	\begin{center}
			\begin{tabular}{l c c c c c c c c }
				\hline
				\hline
				\multicolumn{9}{c}{Scalar perturbations}\\ \hline
				&&\multicolumn{3}{c}{$\omega_0=-2/3$}
				&\multicolumn{4}{c}{$\omega_0=-1/2$}\\
				\hline
				&&\multicolumn{1}{c}{WKB}
				&&\multicolumn{1}{c}{AIM}
				&&\multicolumn{1}{c}{WKB}
				&&\multicolumn{1}{c}{AIM}
				\\ 
				$\rho_0$&\multicolumn{8}{c}{$l=0$}
				\\
				0.01
				&&0.195563-0.180831$i$	
				&&0.195457-0.187169$i$	
				&&0.195975-0.179954$i$
				&&0.195743-0.186603$i$
				\\
				0.10
			&&0.166671-0.174322$i$	
			&&0.167064-0.176485$i$	
			&&0.170528-0.166209$i$
			&&0.170589-0.170975$i$
				\\		
				$\rho_0$&\multicolumn{8}{c}{$l=1$}
				\\
				0.01
				&&0.547237-0.173928$i$	
				&&0.547278-0.173778$i$	
				&&0.547896-0.173768$i$
				&&0.547937-0.173619$i$
				\\
				0.10
				&&0.504477-0.157630$i$	
				&&0.504505-0.157515$i$	
				&&0.510936-0.156189$i$
				&&0.510967-0.156078$i$
				\\			%
				&\multicolumn{8}{c}{$l=3$}\\
				0.01
				&&1.270480-0.171755$i$	
				&&1.270480-0.171753$i$	
				&&1.271550-0.171696$i$
				&&1.271550-0.171695$i$
				\\
				0.10
				&&1.184560-0.154697$i$	
				&&1.184560-0.154696$i$	
				&&1.195430-0.154106$i$
				&&1.195430-0.154105$i$
				\\
				&\multicolumn{8}{c}{$l=5$}\\
				0.01
				&&1.995190-0.171473$i$	
				&&1.995190-0.171473$i$	
				&&1.996760-0.171429$i$
				&&1.996760-0.171429$i$
				\\
				0.10
				&&1.863150-0.154311$i$	
				&&1.863150-0.154311$i$	
				&&1.879280-0.153837$i$
				&&1.879280-0.153837$i$
				\\
				&\multicolumn{8}{c}{$l=7$}\\
				0.01
				&&2.720170-0.171385$i$	
				&&2.720170-0.171385$i$	
				&&2.722270-0.171345$i$
				&&2.722270-0.171345$i$
				\\
				0.10
				&&2.541410-0.154190$i$	
				&&2.541410-0.154190$i$	
				&&2.562970-0.153753$i$
				&&2.562970-0.153753$i$
				\\
				\hline
				\hline
			\end{tabular}
		\caption{Quasinormal frequencies for the Scalar perturbations for several values of the parameter $\rho_0$, with $n=0$ and $\epsilon^2=0.1$.}
		\label{s0r}
	\end{center}
\end{table}

\begin{table}[th]
	\begin{center}
			\begin{tabular}{l c c c c c c c c}
				\hline
				\hline
				\multicolumn{9}{c}{Electromagnetic perturbations}\\ 
				&&\multicolumn{3}{c}{$\omega_0=-2/3$}
				&\multicolumn{4}{c}{$\omega_0=-1/2$}
				\\
				\hline
				&&\multicolumn{1}{c}{WKB}
				&&\multicolumn{1}{c}{AIM}
				&&\multicolumn{1}{c}{WKB}
				&&\multicolumn{1}{c}{AIM}
				\\
				%
				$\epsilon^2$
				&\multicolumn{8}{c}{$l=1$}\\
				0.01				
				&&0.465629-0.165506 $i$	
				&&0.465727-0.165296$i$	
				&&0.469719-0.165173$i$
				&&0.469814-0.164968$i$
				\\
				0.10				
				&&0.444590-0.149292 $i$	
				&&0.444653-0.149151$i$	
				&&0.448856-0.148948$i$
				&&0.448918-0.148810$i$
				\\
				&\multicolumn{8}{c}{$l=3$}\\
				0.01
				&&1.223820-0.170371$i$	
				&&1.223820-0.170370$i$	
				&&1.233200-0.169992$i$
				&&1.233200-0.169991$i$
				\\
				0.10
				&&1.160040-0.153247$i$	
				&&1.160040-0.153246$i$	
				&&1.169890-0.152861$i$
				&&1.169890-0.152860$i$
				\\
				&\multicolumn{8}{c}{$l=5$}\\
				0.01
				&&1.950800-0.170964$i$	
				&&1.950800-0.170964$i$	
				&&1.965510-0.170580$i$
				&&1.965510-0.170580$i$
				\\
				0.10
				&&1.847640-0.153727$i$	
				&&1.847640-0.153727$i$	
				&&1.863110-0.153337$i$
				&&1.879280-0.153837$i$
				\\
				&\multicolumn{8}{c}{$l=7$}\\
				0.01
				&&2.671970-0.171148$i$	
				&&2.671970-0.171148$i$	
				&&2.692020-0.170762$i$
				&&2.692020-0.170762$i$
				\\
				0.10
				&&2.530050-0.153876$i$	
				&&2.530050-0.153876$i$	
				&&2.551140-0.153484$i$
				&&2.551140-0.153484$i$
				\\
				\hline
				\hline
			\end{tabular}
		\caption{Quasinormal frequencies for the Electromagnetic perturbations for several values of the parameter $\epsilon$, with $n=0$ and $\rho_0=0.1$.}
		\label{s1e}
	\end{center}
\end{table}

\begin{table}[th]
	\begin{center}
			\begin{tabular}{l c c c c c c c c}
				\hline
				\hline
				\multicolumn{9}{c}{Electromagnetic perturbations}\\ 
				&&\multicolumn{3}{c}{$\omega_0=-2/3$}
				&\multicolumn{4}{c}{$\omega_0=-1/2$}
				\\
				\hline
				&&\multicolumn{1}{c}{WKB}
				&&\multicolumn{1}{c}{AIM}
				&&\multicolumn{1}{c}{WKB}
				&&\multicolumn{1}{c}{AIM}
				\\
				%
				$\rho_0$
				&\multicolumn{8}{c}{$l=1$}\\
				0.01				
				&&0.472513-0.165563$i$	
				&&0.472604-0.165367$i$	
				&&0.472932-0.165534 $i$
				&&0.473022-0.165339$i$
				\\
				0.10				
			&&0.444590-0.149292$i$	
			&&0.444653-0.149151$i$	
			&&0.448856-0.148948$i$
			&&0.448918-0.148810$i$
				\\
				&\multicolumn{8}{c}{$l=3$}\\
				0.01
			&&1.239670-0.170316$i$	
			&&1.239670-0.170314$i$	
			&&1.240630-0.170282$i$
			&&1.240630-0.170281$i$
				\\
				0.10
			&&1.160040-0.153247$i$	
			&&1.160040-0.153246$i$	
			&&1.169890-0.152861$i$
			&&1.169890-0.152860$i$
				\\
				&\multicolumn{8}{c}{$l=5$}\\
				0.01
				&&1.975680-0.170895$i$	
				&&1.975680-0.170895$i$	
				&&1.977180-0.170861 $i$
				&&1.977180-0.170861$i$
				\\
				0.10
				&&1.847640-0.153727$i$	
				&&1.847640-0.153727$i$	
				&&1.863110-0.153337$i$
				&&1.863110-0.153337$i$
				\\
				&\multicolumn{8}{c}{$l=7$}\\
				0.01
				&&2.705880-0.171074$i$	
				&&2.705880-0.171074$i$	
				&&2.707920-0.171041$i$
				&&2.707920-0.171041$i$
				\\
				0.10
				&&2.530050-0.153876$i$	
				&&2.530050-0.153876$i$	
				&&2.551140-0.153484$i$
				&&2.551140-0.153484$i$
				\\
				\hline
				\hline
			\end{tabular}
		\caption{Quasinormal frequencies for the Electromagnetic perturbations for several values of the parameter $\rho_0$, with $n=0$ and $\epsilon^2=0.1$.}
		\label{s1r}
	\end{center}
\end{table}

In Table \ref{s0e} and Table \ref{s1e}, we can observe the frequencies of the QNM for the scalar and electromagnetic perturbations, the real part increases when $l$ increases and we fix the parameter $\rho_{0}$. The same behavior occurs when the parameter $\epsilon$ is fixed (see Tables  \ref{s0r} and \ref{s1r} ). In the cases of the imaginary part when the parameters $\rho_{0}$ or $\epsilon$ are fixed, their values increases as $l$ increases, for both perturbations, then we can mention that the relaxation time of the Schwdm BH diminishes and the BH is stable. 

An interesting feature, we noticed is that due to quintessence-like matter, the scalar and electromagnetic fields damp more rapidly when the parameter $\omega_0$ increases. Tables I-IV show as the parameter $\omega_0$ increases, the magnitudes of the real part of the frequencies of the QNM increase. In contrast, the absolute value of the imaginary part of the frequencies of the QNM decreases. On the other hand, comparing our results presented in this work with those that use the Kiselev model, the scalar field also decays more rapidly for higher values of state parameter in the case of Schwarzschild surrounded by quintessence, as shown in Ref. \cite{Chen:2005qh}. In this reference, the state parameter plays the same role as the parameter $\omega_0$ in our case.

It can be seen that the range of real and imaginary frequencies is approximately the same and it does not change drastically when the WKB or the AIM are considered, so the AIM is very reliable because in this case few iterations are required to compute the quasinormal modes. On the other hand, the calculation time is much less compared to WKB. This feature can be useful to study another more complicated black hole solution.

\section{Time evolution of Scalar and Electromagnetic perturbations}
\label{sec.4}
In this section we compute the time evolution of the Scalar and Electromagnetic perturbations to further reveal the instability of the Schwarzschild black hole with quintessence-like matter and a deficit solid angle. For the time evolution, the radial part of the perturbation equations are reduced to the form
\begin{equation}\label{ec.ecper}
	\frac{\partial^2\Psi}{\partial t^2}-\frac{\partial^2\Psi}{\partial r_*^2}+V(r)\Psi=0\,.
\end{equation}
After recasting the wave equation (\ref{ec.ecper}) in the null coordinates $u=t-r_*$ and $v=t+r_*$ we obtain
\begin{equation}\label{key}
-4\frac{\partial^2}{\partial u \partial v}\Psi(u,v)=V(u,v)\Psi(u,v)\,.
\end{equation}  
In order to compute the time evolution of $\Psi$ we implemented the discretization scheme developed in \cite{Gundlach:1993tp}, i.e. we can numerically integrate by using the finite difference method. Using the Taylor expansion, one find
\begin{eqnarray}
\Psi(u+\Delta u,v+\Delta v)&=&\Psi(u,v+\Delta v)+\Psi(u+\Delta u,v)-\Psi(u,v)\nonumber\\
&&-\frac{\Delta u\Delta v}{8}V\left(\frac{2v-2u+\Delta v-\Delta u}{4}\right)\left[\Psi(u+\Delta u,v)+\Psi(u,v+\Delta v)\right]+O\left(h^4\right)\,,
\end{eqnarray}
where $h$ is an overall grid scalar factor ($\Delta  u=\Delta v=h$). To perform the numerical integration on an uniformly spaced grid.  we impose the following initial profile 
\begin{eqnarray}\label{key}
\Psi\left(u,v=0\right)=0,\qquad 
\psi\left(u=0,v\right)=Ae^{
-\frac{\left(v-v_0\right)^2}{2\sigma^2}
}\,,
\end{eqnarray} 
since the late-time behavior of the wave function is found to be insensitive to the initial data, we set the initial Gaussian distribution with width, $\sigma=3$ centered at $v_0 = 10$, amplitude 
$A = 1$ and with an overall grid scale factor $h=0.2$. To proceed the integration in the aforementioned scheme one has to find the value of the potential at $r(r_*)$ at each step. 

For the special case $\omega_0=-2/3$, we get
\begin{equation}\label{key}
r_*=\frac{1}{2\kappa_+}\ln\left|\frac{r}{r_+}-1\right|	+\frac{1}{2\kappa_c}\ln\left|1-\frac{r}{r_c}\right|	\,,
\end{equation}
and for $\omega_0=-1/2$, we obtain
\begin{equation}\label{key}
	r_*=-\frac{3\sqrt{r}}{\rho_0}-\frac{1}{2\kappa_+}\ln\left|\frac{\sqrt{r}}{r_+}-1\right|	-\frac{3r_1^3}{\rho_0\left(r_1-1\right)\left(r_1-r_2\right)}\ln\left|1-\frac{\sqrt{r}}{r_1}\right|	
	+\frac{3r_2^3}{\rho_0\left(r_2-1\right)\left(r_1-r_2\right)}\ln\left|1-\frac{\sqrt{r}}{r_2}\right|\,,
\end{equation}
where
\begin{equation}\label{key}
r_{1,2}=\frac{1}{4\rho_0}\left[
3\left(1-\epsilon^2\right)-2\rho_0\pm
\sqrt{
3\left[3\left(1-\epsilon^2\right)-2\rho_0\right]
\left[1-\epsilon^2+2\rho_0\right]}
\right]\,.
\end{equation}
So we get $r(r_*)$ numerically, using the built-in Mathematica commands. 

\begin{figure}[!h]
	\centering
	\includegraphics[scale=0.96]{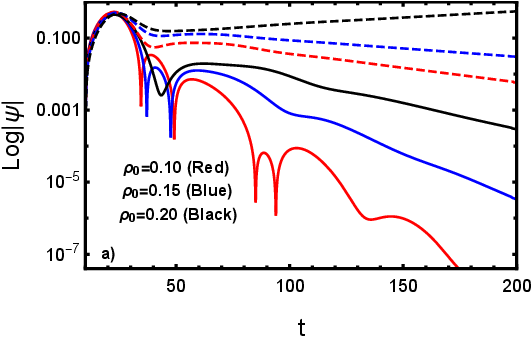}
	\includegraphics[scale=0.96]{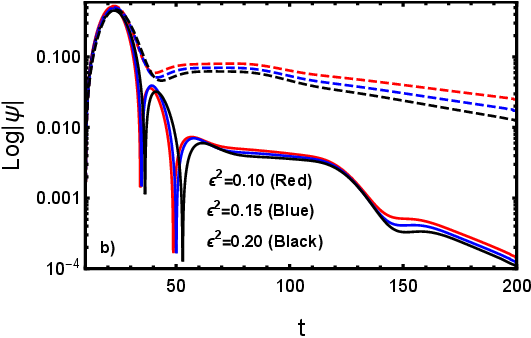}
	\caption{Logarithmic plot of the evolution of a scalar field in Schwdm BH for the case of $l=0$ mode. a) the plot correspond to  $\epsilon=0.1$ and b) the plot correspond to $\rho_0=0.1$. In both graphs, the solid line is for  $\omega_0=-1/2$, while the dotted line is for $\omega_0=-2/3$.}
	\label{g7}
\end{figure}

In Fig. \ref{g7} we show the evolution profile of the scalar field for $l=0$ mode. We observe that the $\mathrm{Log}\left|\Psi\right|$ linearly depends on $t$ at later times. From Fig. \ref{g7}(a) we see that the perturbations start to become unstable for higher values of $\rho_0$, whereas if the values of $\epsilon^2$ increases, the system becomes stable. In both graphs when the value of the quintessence parameter $\omega_0$ increases, the system becomes more stable, which is in correspondence with the peculiar shape of the potential for $\omega_0=-2/3$ and $\omega_0=-1/2$ (Fig. \ref{g4}). 

\begin{figure}[!h]
	\centering
	\includegraphics[scale=0.96]{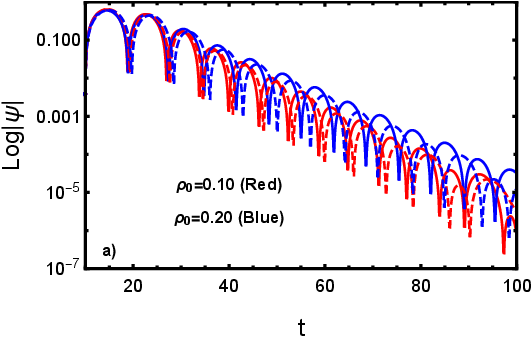}
	\includegraphics[scale=0.96]{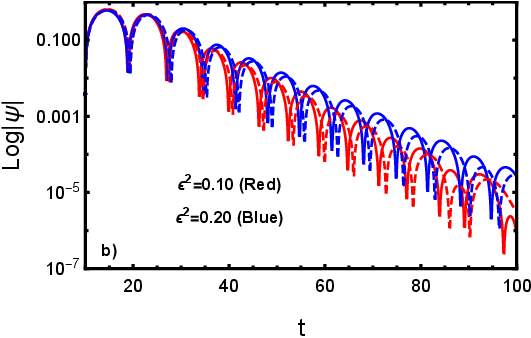}
	\caption{Logarithmic plot of the evolution of a scalar field in Schwdm BH for the case of $l=1$ mode. a) the plot correspond to  $\epsilon=0.1$ and b) the plot correspond to $\rho_0=0.1$. In both graphs, the solid line is for  $\omega_0=-1/2$, while the dotted line is for $\omega_0=-2/3$.}
	\label{g8}
\end{figure} 

\begin{figure}[!h]
	\centering
	\includegraphics[scale=0.96]{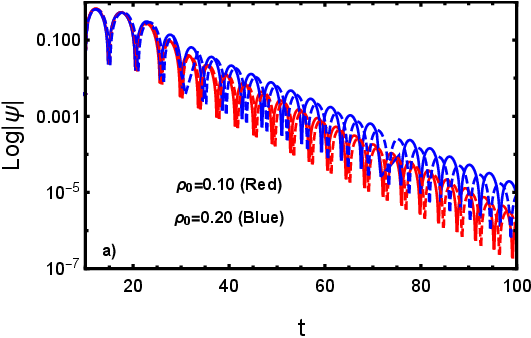}
	\includegraphics[scale=0.96]{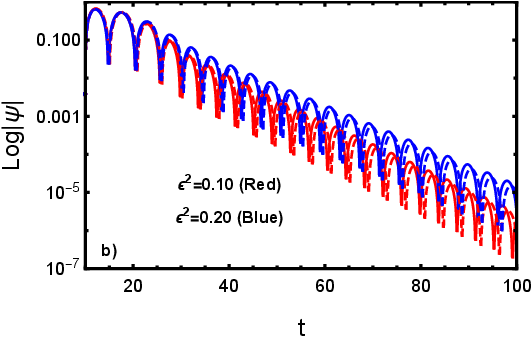}
	\caption{Logarithmic plot of the evolution of a electromagnetic field in Schwdm BH for the case of $l=2$ mode. a) the plot correspond to  $\epsilon=0.1$ and b) the plot correspond to $\rho_0=0.1$. In both graphs, the solid line is for  $\omega_0=-1/2$, while the dotted line is for $\omega_0=-2/3$.}
	\label{g9}
\end{figure} 

Fig. \ref{g8} demonstrates the evolution of scalar field around of the Schwdm BH with different $\rho_0$, $\epsilon^2$ and $\omega_0$, and in the Fig. \ref{g9} the evolution of electromagnetic field is showed.

One can see from Fig. \ref{g8} and Fig. \ref{g9}  that difference of the temporal evolution of the perturbations in the Schwdm BH, is that increases the frequencies of oscillation in the case of the electromagnetic perturbation. Also, it is possible to notice that at $\omega_{0}=-1/2$ and $\omega_{0}=-2/3$ of the Schwdm BH, the evolution of the perturbations are almost the same. Moreover, time domain profiles of the scalar and electromagnetic perturbations of the black hole show that it is stable against perturbations.

\section{Conclusions}\label{sec.5}

In this contribution, we have focused on scalar and electromagnetic perturbations of spherically symmetric Schwarzschild black hole with quintessence-like matter and a deficit solid angle. First, we numerically calculated the real and imaginary frequencies of the QNM. Secondly, we studied the time evaluation of scalar and electromagnetic perturbations.    

We calculated the QNM of Schwdm BH, we apply mainly the improved AIM and WKB approximation sixth order to make a comparison between both methods. The magnitudes of the real part and the imaginary part of the QNM for scalar and electromagnetic perturbations decrease with the decreases of values of the density of quintessence-like matter and deficit solid angle. The  frequencies are approximately the same and it does not change drastically when the WKB method or the AIM are considered, so AIM is very reliable, in this particular case, because few iterations are required to compute the quasinormal modes also the calculation time is much less compared to WKB.   

From Tables \ref{s0e}-\ref{s1r}, we can see that the change with respect to $\omega_0$ of the real part and the magnitude of the imaginary part of the frequencies is very small, while the real part increase with increasing $l$ and the increase of the magnitude of imaginary part of the frequencies is very small.

The evolution of scalar and electromagnetic perturbations obtained in this work confirms that in the QNM stage, all the fields decay slowly due to the presence of quintessence-like matter. At late times, the frequencies are suppressed by the tail form of field decay. However, from Fig. \ref{g7}(a), we can see that the late decay of scalar perturbations with $l = 0$ gives up the power law form of decay, relaxing to a constant field. This constant value of the field increases as the values of $\rho_0$ increase, while it decreases if the values of $\epsilon$ increase (This behavior can be understood as a consequence of cosmological no-hair theorem).

We also saw the behavior of perturbations with changing parameter $\rho_0$, $\epsilon^2$ and $\omega_0$. We note that perturbations with the higher value of $\rho_0$ becomes unstable for the $l=0$ mode of the scalar field. The behavior of the perturbation for different values of the  $\epsilon^2$ and $\omega_0$ is also studied and reported.

In general, we also observed that the different parameters have a critical role in black hole stability. The results obtained for QNM frequencies can play a significant role as an indirect way of detecting dark energy near the black hole by using future gravitational wave detectors for interactions of scalar fields with the black hole.

\appendix*
\section{Improved AIM}

In this appendix we briefly explain the improved AIM. We shall start with the second-order
differential equation for the function $\chi(\xi)$ 
\begin{equation}\label{ec.aim1}
	\chi''=\lambda_0(\xi)\chi'+s_0(\xi)\chi\,.	
\end{equation} 
We can express higher derivatives of $\chi(\xi)$ in terms of $\chi'(\xi)$ and $\chi(\xi)$, as following 
\begin{equation}\label{ec.aim2}
	\chi^{(n+2)}=\lambda_n(\xi)\chi'+s_n(\xi)\chi\,,
\end{equation}
where
\begin{eqnarray}
	\lambda_n(\xi)&=&\lambda'_{n-1}(\xi)+s_{n-1}(\xi)+\lambda_0(\xi)\lambda_{n-1}(\xi)\,,\label{ec.ln0}\\
	s_n(\xi)&=&s'_{n-1}(\xi)+s_0(\xi)\lambda_{n-1}(\xi)\,.\label{ec.sn0}
\end{eqnarray}
For sufficiently large $n$, the
coefficients $\lambda_n(\xi)$ and $s_n(\xi)$ satisfy the quantization relation
\begin{equation}\label{ec.qz}
	s_n\lambda_{n-1}-s_{n-1}\lambda_n=0\,.	
\end{equation}
Here, the eigenvalues are given by roots of (\ref{ec.qz}). However, for each iteration, we must take derivatives of $\lambda_{n-1}$ and $s_{n-1}$ to compute $\lambda_{n}$ and $s_{n}$, so this process can slow the numerical implementation of the AIM down considerably and it can increase execution time. To overcome this issues, an improved version of AIM was developed in Ref. \cite{Cho:2009cj}, which avoids the need to take derivatives at each step.

Now, expand the $\lambda_n(\xi)$ and $s_n(\xi)$ in a Taylor series around the point $\xi_0$, we have
\begin{eqnarray}
	\lambda_n(\xi)&=&\sum_{i=0}^{\infty}c_n^i\left(\xi-\xi_0\right)^i\,,\label{ec.ln}\\
	s_n(\xi)&=&\sum_{i=0}^{\infty}d_n^i\left(\xi-\xi_0\right)^i\,,\label{ec.sn}
\end{eqnarray} 
where $c_n^i$ and $d_n^i$ are the $i$th Taylor coefficients. Substituting (\ref{ec.ln}) and (\ref{ec.sn}) in (\ref{ec.ln0}) and (\ref{ec.sn0}), we have the following recursion relations for the coefficients
\begin{eqnarray}
	c_n^i&=&(i+1)c_{n-1}^{i+1}+d_{n-1}^i+\sum_{j=0}^ic_0^jc_{n-1}^{i-j}\,,\label{ec.ln1}\\
	d_n^i&=&(i+1)d_{n-1}^{i+1}+\sum_{j=0}^id_0^jc_{n-1}^{i-j}\,,\label{ec.sn1}
\end{eqnarray} 
then, the quantization condition (\ref{ec.qz}) can be rewritten as
\begin{equation}\label{ec.qz1}
	d_n^0c_{n-1}^0-d_{n-1}^0c_n^0=0\,.
\end{equation}
Using the quantization condition (\ref{ec.qz1}) we can obtain QNM, for $n$ large enough.

\section*{ACKNOWLEDGMENT}
The authors acknowledge the partial financial support from SNII--CONAHCYT, M\'exico.

\bibliographystyle{unsrt}
\bibliography{bibliografia}

\end{document}